\documentclass[a4paper,11pt]{amsart}

\usepackage{amssymb}
\usepackage{mhequ}
\usepackage{cite}

\newcommand{\cQ}{\mathcal Q}
\newcommand{\cZ}{\mathcal Z}
\newcommand{\cL}{\mathcal L}
\DeclareFontFamily{OT1}{rsfs}{}
\DeclareFontShape{OT1}{rsfs}{m}{n}{ <-7> rsfs5 <7-10> rsfs7 <10->
rsfs10}{} \DeclareMathAlphabet{\mathscr}{OT1}{rsfs}{m}{n}

\newcommand{\mcM}{\mathscr M}

\newcommand{\eq}[1]{\eqref{#1}}

\newcommand{\bel}[1]{\begin{equation}\label{#1}}
\newcommand{\beal}[1]{\begin{eqnarray}\label{#1}}
\newcommand{\beadl}[1]{\begin{deqarr}\label{#1}}
\newcommand{\eeadl}[1]{\arrlabel{#1}\end{deqarr}}
\newcommand{\eeal}[1]{\label{#1}\end{eqnarray}}
\newcommand{\eead}[1]{\end{deqarr}}
\newcommand{\eea}{\end{eqnarray}}
\newcommand{\eeaa}{\end{eqnarray*}}

\newcommand{\be}{\begin{equation}}
\newcommand{\ee}{\end{equation}}

\DeclareFontFamily{OT1}{rsfs}{}
\DeclareFontShape{OT1}{rsfs}{m}{n}{ <-7> rsfs5 <7-10> rsfs7 <10->
rsfs10}{} \DeclareMathAlphabet{\mycal}{OT1}{rsfs}{m}{n}

\newcommand{\mcL}{{\mycal L}}
\newcounter{mnotecount}[section]

\newcommand{\N}{{\Bbb N}}

\newcommand{\rmnote}[1]{}

\newcommand{\Ric}{\operatorname{Ric}}

\def\mysavedown#1{\edef\mysubs{\mysubs#1}}
\def\mysaveup#1{\edef\mysups{\mysups#1}}
\def\mydown#1{{\mytensor}_{\vphantom{\mysubs}#1}}
\def\myup#1{{\mytensor}^{\vphantom{\mysups}#1}}
\def\tensor#1#2{
  #1
  \def\mytensor{\vphantom{#1}}
  \def\mysubs{\relax}
  \def\mysups{\relax}
  \let\down=\mysavedown
  \let\up=\mysaveup
  #2
  \let\down=\mydown
  \let\up=\myup
  #2
  }

\newcommand{\Hess}{\operatorname{Hess}}

\newcommand{\Tr}{\operatorname{Tr}}
\newcommand{\grav}{\operatorname{grav}}
\newcommand{\Id}{\operatorname{Id}}

\newcommand{\R}{\mathbb R}

\newcommand{\mbbS}{\mathbb S}

\renewcommand{\setminus}{\smallsetminus}

\renewcommand{\to}{\rightarrow}
\renewcommand{\div}{\operatorname{div}}

\newcommand{\cross}{\mathbin{\times}}

\renewcommand{\epsilon}{\varepsilon}
\renewcommand{\hat}{\widehat}
\let\scr=\mathscr

\def\crn#1#2{{\vcenter{\vbox{
        \hbox{\kern#2pt \vrule width.#2pt height#1pt
           }
          \hrule height.#2pt}}}}

\newcommand{\del}{\partial}
\newcommand{\newF}{\lambda}

\renewcommand{\hbar}{{\overline h}}

\newcommand{\w}{\widetilde}

\newcommand{\pre}[2]{{{\vphantom{#2}}^{#1}}\kern-.2ex{#2}}

\theoremstyle{plain}
\newtheorem{theorem}{\sc Theorem}[section]
\newtheorem{lemma}[theorem] {\sc Lemma}
\newtheorem{proposition}[theorem]{\sc Proposition}
\newtheorem{corollary}[theorem] {\sc Corollary}

\theoremstyle{definition}

\newtheorem{remark}[theorem]{\sc  Remark\rm}

\numberwithin{equation}{section}

\date{\today}

\begin{document}

\title[Stationary space-times with negative $\Lambda$] {Non-singular,
vacuum, stationary space-times with a negative cosmological constant}
\author[P.T. Chru\'sciel]{Piotr T.~Chru\'sciel} \address{Piotr
T.~Chru\'sciel, LMPT, F\'ed\'eration Denis Poisson, Facult\'e des
Sciences, Parc de Grandmont, F37200 Tours, France. }
\email{piotr.chrusciel@lmpt.univ-tours.fr} \urladdr{
http://www.phys.univ-tours.fr$/\sim${piotr}} \author[E.  Delay]{Erwann
Delay} \address{Erwann Delay, Laboratoire d'analyse non lin\'eaire et
g\'eom\'etrie, Facult\'e des Sciences, 33 rue Louis Pasteur, 84000
Avignon, France} \email{Erwann.Delay@univ-avignon.fr}
\urladdr{http://www.math.univ-avignon.fr/Delay}
\begin{abstract}
We construct infinite dimensional families of non-singular stationary
space times, solutions of the vacuum Einstein equations with a negative cosmological constant.
\end{abstract}

\maketitle

\tableofcontents
\section{Introduction}\label{section:intro}

 A class of space-times of  interest is
that of vacuum metrics with a negative cosmological constant admitting
a smooth conformal completion at infinity. It is natural to seek for
stationary solutions with this property. In this paper we show that a
large class of such solutions can be constructed by prescribing the
conformal class of a stationary Lorentzian metric on the conformal
boundary $\partial \mcM$, provided that the boundary data are
sufficiently close to, e.g., those of anti-de Sitter space-time.

We mention the recent papers~\cite{ACD,ACD2}, where we have
constructed infinite dimensional families of \emph{static},
singularity free solutions of the vacuum Einstein equations with a
negative cosmological constant. The main point of the current work is
to remove the staticity restriction. This leads to new, infinite
dimensional families of non-singular, \emph{stationary} solutions of
those equations.

We thus seek to construct Lorentzian metrics ${}^{n+1}g$ in any
space-dimension $n\geq 2$, with Killing vector
$X=\partial/\partial t$. In adapted coordinates those metrics can
be written as \beal{gme1} &^{n+1}g = -V^2(dt+\underbrace{\theta_i
dx^i}_{=\theta})^2 + \underbrace{g_{ij}dx^i dx^j}_{=g}\;, & \\ &
\partial_t V = \partial_t \theta = \partial_t g=0\;.\eeal{gme2}

Our main result reads as follows (see below for the definition of
non-degeneracy; the function $\rho$ in \eq{result} is a coordinate
near $\partial M$ that vanishes at $\partial M$):

\begin{theorem}\label{maintheorem}
 Let $n=\dim M\ge 2$, $k\in\N\setminus\{0\}$, $\alpha\in(0,1)$, and
 consider a static Lorentzian Einstein metric of the form
 \eq{gme1}-\eq{gme2} with strictly positive $V=\mathring V$,
 $g=\mathring g$, and $\theta=0$, such that the associated Riemannian
 metric $\w{g}={\mathring V}^2d\varphi^2+\mathring g$ on
 $\mbbS^1\times M$ is $C^2$ compactifiable and non-degenerate, with
 smooth conformal infinity.  For every smooth $\hat{\theta}$,
 sufficiently close to zero in $C^{k+2,\alpha}(\partial M,{\mathcal
 T}_1)$, there exists a unique, modulo diffeomorphisms which are the
 identity at the boundary, nearby stationary vacuum metric of the form
 \eq{gme1}-\eq{gme2} such that, in local coordinates near the
 conformal boundary $\partial M$, \bel{result} V-\mathring
 V=O(\rho)\;,\quad \theta_i=\hat \theta_i +O(\rho) ,\quad
 g_{ij}-\mathring g_{ij} =O(1)\;.\ee
\end{theorem}

Theorem~\ref{maintheorem} is more or less a rewording of
Theorem~\ref{imp} below, taking into account the discussion of
uniqueness in Section~\ref{SecUni}.

The $(n+1)$-dimensional anti-de Sitter metric is non-degenerate in the
sense above, so Theorem~\ref{maintheorem} provides in particular an
infinite dimensional family of solutions near that metric.

The requirement of strict positivity of $\mathring V$ excludes black
hole solutions, it would be of interest to remove this condition.

The decay rates in \eq{result} have to be compared with the
leading order behavior $\rho^{-2}$ both for ${\mathring V}^2$ and
${\mathring g}_{ij}$. A precise version of \eq{result} in terms of
weighted function spaces (as defined below) reads
\beal{result2}
 & (V-\mathring V)\in
C_{1}^{k+2,\alpha}(\mbbS^1\times M)\;,\;\;(g-\mathring g)\in
C_{2}^{k+2,\alpha}(\mbbS^1\times M,{\mathcal S}_2)\;,
 & \\ &
 \quad
\theta-\hat \theta \in C_{2}^{k+2,\alpha}(\mbbS^1\times M,
{\mathcal T}_1)\;,
 &  \eeal{result3}
 and the
norms of the differences above are small in those spaces.

Note that our hypothesis that the metric $\w{g}$ is conformally
$C^2$ implies that $\w{g} $ is $C^{n-1,\alpha}\cap
C^{3,\alpha}$--conformally compactifiable and
polyhomogeneous~\cite{CDLS}. We show in
Section~\ref{section:polyhomogeneity} that our solutions have
complete polyhomogeneous expansions near the conformal boundary,
see Theorem~\ref{Tphg} for a precise statement. Since the
Fefferman-Graham expansions are valid regardless of the signature
of the boundary metric, the solutions are smooth in even
space-time dimensions. In odd space-time dimensions the
obstruction to smoothness is the non-vanishing of the
Fefferman-Graham obstruction tensor~\cite{FG,GrahamHirachi} of the
(Lorentzian) metric obtained by restricting
$-(dt+\theta)^2+V^{-2}g $ to the conformal boundary at infinity.

Theorem~\ref{maintheorem} is proved by an implicit-function
argument. This requires the proof of isomorphism properties of an
associated linearised operator. This operator turns out to be
rather complicated, its mapping properties being far from evident.
We overcome this by reinterpreting this operator as the
Lichnerowicz operator $\tilde\Delta_L+2n$ in one-dimension higher.
Our non-degeneracy condition above is then precisely the condition
that $\tilde\Delta_L+2n$ has no $L^2$--kernel. While this is
certainly a restrictive condition, large classes of Einstein
metrics satisfying this condition are
known~\cite{Lee:fredholm,ACD2,mand1,mand2}.


\section{Definitions, notations and conventions}\label{sec:def}

Let $\overline{N}$ be a smooth, compact $(n+1)$-dimensional manifold
with boundary $\partial {N}$. Set
$N:=\overline{N}\backslash\partial{N}$, $N$ is thus a non-compact
manifold without boundary. In our context the boundary $\partial
{N}$ will play the role of a \emph{boundary at infinity} of $N$. Let
$g$ be a Riemannian metric on $N$, we say that $(N,g)$ is {\it
conformally compact} if there exists on $\overline{N}$ a smooth
defining function $\rho$ for $\partial N$ (that is $\rho\in
C^\infty(\overline{N})$, $\rho>0$ on $N$, $\rho=0$ on $\partial {N}$
and $d\rho$ nowhere vanishing on $\partial N$) such that
$\overline{g}:=\rho^{2}g$ is a $C^{2,\alpha}(\overline {N})\cap
C^{\infty}_0(N)$ Riemannian metric on $\overline{N}$, we will denote
by $\hat{g}$ the metric induced on $\partial N$. Our definitions of
function spaces follow~\cite{Lee:fredholm}. Now if
$|d\rho|_{\overline{g}}=1$ on $\partial N$, it is well known (see
\cite{Mazzeo:hodge} for instance) that $g$ has asymptotically
sectional curvature $-1$ near its boundary at infinity, in that case
we say that $(N,g)$ is {\it asymptotically hyperbolic}. If we assume
moreover than $(N,g)$ is Einstein, then asymptotic hyperbolicity
enforces the normalisation \bel{EE} \Ric(g)=-n g\;, \ee where
$\Ric(g)$ is the Ricci curvature of $g$.

We recall that the  Lichnerowicz Laplacian acting on a symmetric
two-tensor field is defined as~\cite[\S~1.143]{besse:einstein}
$$
\Delta_Lh_{ij}=-\nabla^k\nabla_kh_{ij}+R_{ik}h^k{_j}+R_{jk}h^k{_j}-2R_{ikjl}h^{kl}\;.
$$
The operator $\Delta_L+2n$ arises naturally when  linearising \eq{EE}.
We will say that $g$ is {\it non-degenerate} if $\Delta_L+2n$ has no
$L^2$-kernel.

While we seek to construct metrics of the form \eq{gme1}, for the
purpose of the proofs we will often work with manifolds $N$ of the
form $$ N=\mbbS^1\times M, $$ equipped with a warped product,
asymptotically hyperbolic metric $$V^2d\varphi^2+g,$$ where $V$ is a
positive function on $M$ and $g$ is a Riemannian metric on $M$. By an
abuse of terminology, such metrics will be said {\it static}.


The basic example of a non-degenerate, asymptotically hyperbolic,
static Einstein space is the Riemannian counterpart of the AdS
space-time. In that case $M$ is the unit ball of $\R^n$, with the
hyperbolic metric $$g_0=\rho^{-2}\delta\;,$$ $\delta$ is the Euclidean
metric, $\rho(x)=\frac{1}{2}(1-|x|_\delta^2)$, and
$$V_0=\rho^{-1}-1\;.$$

We denote by ${\mathcal T}^q_p$ the set of rank $p$ covariant and rank
$q$ contravariant tensors. When $p=2$ and $q=0$, we denote by
${\mathcal S}_2$ the subset of symmetric tensors. We use the summation
convention, indices are lowered and raised with $g_{ij}$ and its
inverse $g^{ij}$.

\section{Isomorphism theorems}
 Some of the isomorphism theorems we will use are consequences of
 Lee's theorems \cite{Lee:fredholm}, it is therefore convenient to
 follow his notation for the weighted H\"older spaces
 $C^{k,\alpha}_\delta$. As described in the second paragraph before
 proposition B of \cite{Lee:fredholm}, a tensor in this space
 corresponds to $\rho^\delta$ times a tensor in the usual
 $C^{k,\alpha}$ space as defined  using the norm of the conformally
 compact metric. This implies that, in local coordinates near the
 conformal boundary, a function in $C^{k,\alpha}_\delta$ is
 $O(\rho^\delta)$, a one-form in $C^{k,\alpha}_\delta$ has components
 which are $O(\rho^{\delta-1})$, and a covariant two-tensor in
 $C^{k,\alpha}_\delta$ has components which are $O(\rho^{\delta-2})$.

We will use the isomorphism theorems of~\cite{Lee:fredholm} in
weighted $C^{k,\alpha}$ spaces, for $k\in\N$.  Under the regularity
conditions on the metric in our definition of asymptotically
hyperbolic metric, those theorems apparenly only apply to low values
of $k$. However, under our hypotheses, one can use those theorems
for $k=2$, and use scaling estimates to obtain the conclusion for
any value of $k$.

\subsection{An isomorphism on two-tensors}

We first recall a result of Lee (see Theorem C(c) and proposition
D of~\cite{Lee:fredholm}, there is no $L^2$-kernel here by
hypothesis):

\begin{theorem} Let $\mbbS^1\times M$ be
equipped with a non-degenerate asymptotically hyperbolic metric
$\w{g}$. For $k\in\N$, $\alpha\in(0,1)$ and $\delta\in(0,n)$ the
operator $\w{\Delta}_L+2n$ is an isomorphism from
$C^{k+2,\alpha}_\delta(\mbbS^1\times M,{\mathcal S}_2)$ to
$C^{k,\alpha}_\delta(\mbbS^1\times M,{\mathcal S}_2)$.
\end{theorem}

When the metric is static  of the form $\w{g}=V^2d\varphi^2+g$ we
deduce
\begin{corollary}\label{isotordu}
On $(M,g)$ we consider the operator $$ (W,h)\mapsto (l(W,h),L(W,h))\;,
$$ where
\begin{eqnarray*}l(W,h)&=&V\displaystyle{\left[
(\nabla^*\nabla+2n+V^{-1}\nabla^*\nabla V+V^{-2}|dV|^2)
W+V^{-1}\nabla_jV\nabla^jW\right.}\\&&\left.-V^{-1}\nabla^jV\nabla^kVh_{kj}+\langle
\Hess_gV,h\rangle_g \right].\end{eqnarray*} and
\begin{eqnarray*}
L_{ij}(W,h)&=&\frac{1}{2}\Delta_L h_{ij}+nh_{ij}-\frac{1}{2}V^{-1}\nabla^kV\nabla_kh_{ij}\\
&&+\frac{1}{2}V^{-2}(\nabla_iV\nabla^kVh_{kj}+\nabla_jV\nabla^kVh_{ki})\\
&&-\frac{1}{2}V^{-1}(\nabla_i\nabla^kVh_{kj}+\nabla_j\nabla^kVh_{ki})
\\
&&+2V^{-2}W (\Hess_g V)_{ij} -2V^{-3}\nabla_iV\nabla_jV W.\\
\end{eqnarray*}
Then $(l,L)$ is an isomorphism from
$C^{k+2,\alpha}_{\delta-1}(M)\times C^{k+2,\alpha}_{\delta}(
M,{\mathcal S}_2)$ to $C^{k,\alpha}_{\delta-2}(M)\times
C^{k,\alpha}_{\delta}( M,{\mathcal  S}_2)$ when $\delta\in(0,n)$.
\end{corollary}
\begin{proof}
First, it is easy to see that the Laplacian commutes with the Lie
derivative operator in the Killing direction, so the operator
$\w{\Delta}_L+2n$ restricted to $\varphi$-independent tensor field is
again an isomorphism. Now, from Lemma~\ref{lichnetordu} below, if we
define ${\mathcal P}$ to be the set of symmetric covariant two tensors
of the form $$\w{h}=2VWd\varphi^2+h_{ij}dx^idx^j\;,$$ and if we let
${\mathcal T}$ denote the collection of tensors of the form
$$\w{h}=2\xi_idx^id\varphi,$$ then the Lichnerowicz Laplacian
preserves the decomposition ${\mathcal P}\oplus{\mathcal T}$. In
particular the operator $\frac{1}{2}\w{\Delta}_L+n$ restricted to
${\mathcal P}$ is an isomorphism, and this operator is $(l,L)$.
\end{proof}

\subsection{Two isomorphisms on one-forms}

The proof of Corollary~\ref{isotordu} also shows the following (note a
shift in the rates of decay, as compared to the previous section, due
to the fact that a tensor field $\xi_i dx^i d\varphi$ is in
$C^{m,\sigma}_\rho$ if and only if the one-form $\xi_idx^i$ is in
$C^{m,\sigma}_{\rho-1}$):

\begin{corollary}\label{isotrans} The operator on one-forms defined as
\begin{eqnarray*}{\mathcal L}:\xi_i
 &\mapsto& -\nabla^k\nabla_k\xi_i+V^{-1}\nabla^kV\nabla_k \xi_i
+3V^{-2}\nabla_i V\nabla^k V \xi_k\\&&
+R^l{_i}\xi_l-3V^{-1}\nabla_i\nabla_jV\xi^j+2n\xi_i\;,\end{eqnarray*}
is an isomorphism from $C^{k+2,\alpha}_{\delta-1}( M,{\mathcal T}_1)$
to $ C^{k,\alpha}_{\delta-1}( M,{\mathcal T}_1)$ when
$\delta\in(0,n)$. If we let $\xi=V^2\theta$, we therefore obtain that
the operator $\cQ$ on one-forms defined as $V^{-2}\cL(V^2\theta_i)$
\begin{eqnarray*}{\mathcal Q}:\theta_i&\mapsto& -\nabla^k\nabla_k\theta_i-3V^{-1}\nabla^kV\nabla_k
 \theta_i -2V^{-1}\nabla^k\nabla_kV\theta_i
 +3V^{-2}\nabla_i V\nabla^k V
 \theta_k\\&&
 +R^l{}_i\theta_l-3V^{-1}\nabla_i\nabla_jV\theta^j+2n\theta_i\;,\end{eqnarray*}
 is an isomorphism from $C^{k+2,\alpha}_{\delta+1}(
M,{\mathcal T}_1)$ to $ C^{k,\alpha}_{\delta+1}( M,{\mathcal
T}_1)$ when $\delta\in(0,n)$.

\end{corollary}

We will appeal to yet another result of Lee (see
\cite{Lee:fredholm} Theorem C(c), Proposition F and Corollary 7.4,
there is again no $L^2$-kernel here because of the Ricci curvature
condition):
\begin{theorem} On $\mbbS^1\times M$
equipped with an asymptotically hyperbolic metric $\w{g}$ with negative Ricci
 curvature, the operator $\w{\nabla}^*\w{\nabla}-\w{\Ric}$ acting on
 one-forms is an isomorphism from $C^{k+2,\alpha}_{\delta}(\mbbS^1\times
 M,{\mathcal T}_1)$ to $C^{k,\alpha}_{\delta}(\mbbS^1\times M,{\mathcal
 T}_1)$ when $|\delta-\frac{n}{2}|<\sqrt{\frac{n^2}{4}+1}$.
\end{theorem}
When the metric is static  of the form $\w{g}=V^2d\varphi^2+g$ we
deduce:

\begin{corollary}\label{cor:isoform}Under the  hypotheses of the
preceding theorem, on $(M,g)$  consider the operator
$$
\Omega_i \mapsto
{B}(\Omega)_i+R_{ij}\Omega^j-V^{-1}\nabla_i\nabla^jV\Omega_j=:{\mathcal
B}(\Omega)_i\;,
$$
where
$$
{
B}(\Omega)_i:=\nabla^k\nabla_k\Omega_i+V^{-1}\nabla^kV\nabla_k\Omega_i
 -V^{-2}\nabla_iV\nabla^kV\Omega_k\;.
 $$
Then $\mathcal B$ is an isomorphism from $C^{k+2,\alpha}_{\delta}(
M,{\mathcal T}_1)$ to $C^{k,\alpha}_{\delta}(M,{\mathcal T}_1)$
when $|\delta-\frac{n}{2}|<\sqrt{\frac{n^2}{4}+1}$.
\end{corollary}
\begin{proof}
The argument is identical to the proof of Corollary~\ref{isotordu}
using Lemma~\ref{deltaform} and the fact that, in the notation of
Lemma~\ref{deltaform},
$$\w{R}_{ic}\w{\Omega}^c=R_{ij}\Omega^j-V^{-1}\nabla_i\nabla^jV\Omega_j\;.$$
\end{proof}

\subsection{An isomorphism on functions in dimension n}

If we assume that $V^2d\varphi^2+g$ is a static asymptotically
hyperbolic metric on $\mbbS^1\times M$, then it is easy to check that at
infinity $V^{-2}|dV|^2=1$ and $V^{-1}\nabla^i\nabla_iV=n$.  In
dimension $n$, we will need an isomorphism property for the following
operator acting on functions: $$
\sigma\mapsto {\mathcal
T}\sigma:=V^{-3}\nabla^i(V^{3}\nabla_i\sigma)=
\nabla^i\nabla_i\sigma+3V^{-1}\nabla^iV\nabla_i\sigma\;.
$$ From \cite[Theorem~7.2.1 (ii) and Remark (i),
p.~77]{AndChDiss} we obtain:

\begin{theorem}\label{isofunction}
Let $(V,g)$ be close in $C_{-1}^{k+2,\alpha}(M)\times
C_{0}^{k+2,\alpha}(M,{\mathcal S}_2)$ to an asymptotically hyperbolic
static metric. Then $\mathcal T$ is an isomorphism from
$C^{k+2,\alpha}_{\delta}( M)$ to $C^{k,\alpha}_{\delta}(M)$ when
$0<\delta< n+2$.
\end{theorem}

\begin{remark}
Theorem~\ref{isofunction} will be used with $\sigma=O(\rho^2)$,
note that $\delta=2$ verifies the inequality above since $n\ge
2$.
\end{remark}

\subsection{An isomorphism on functions in dimension 3}

 In dimension $n=3$, we will also be interested in the following
 operator acting on functions: $$
\omega\mapsto {\mathcal
Z}\omega:=V^3\nabla^i(V^{-3}\nabla_i\omega)=
\nabla^i\nabla_i\omega-3V^{-1}\nabla^iV\nabla_i\omega\;.
$$ The indicial exponents for this equation are $\mu_-=-1$ and
$\mu_+=0$ (see \cite[Remark (i), p.~77]{AndChDiss}). As $\mu_+\not
> 0$ we cannot invoke \cite[Theorem~7.2.1]{AndChDiss} to
conclude. Instead we appeal to the results of
Lee~\cite{Lee:fredholm}.  For this we need to have a formally
self-adjoint operator, so we set $\omega=V^{\frac{3}{2}}f$, thus
\bel{ZcalZ} {\mathcal
Z}\omega=V^{\frac{3}{2}}\left[\nabla^i\nabla_if-\left(\frac{15}{4}
V^{-2}|dV|^2-\frac{3}{2}V^{-1}\nabla^i\nabla_iV\right)f\right]=:
V^{\frac{3}{2}}Zf\;. \ee At infinity $V^{-2}|dV|^2=1$ and
$V^{-1}\nabla^i\nabla_iV=3$, leading to the following indicial
exponents
 $$ \delta=\frac{1}{2},\frac{3}{2}\;.
$$ We want to show that $Z$ satisfies condition (1.4) of
\cite{Lee:fredholm}, \bel{Lec1.4}\|u\|_{L^2}\le C \|{
Z}u\|_{L^2}\;,\ee for smooth $u$ compactly supported in a
sufficiently small open set ${\mathcal U}\subset M$ such that
$\overline{\mathcal U}$ is a neighborhood of $\partial M$.  We
will need the following, well known result; we give the proof for
completeness:
\begin{lemma}\label{Lnew}
On an asympotically hyperbolic manifold $(M,g)$ with boundary
definining function $\rho$ we have, for all compactly supported $C^2$
functions,
$$
\int u\nabla^*\nabla u\geq
\left(\frac{n-1}{2}\right)^2\int(1+O(\rho)) u^2.
$$
\end{lemma}
\begin{proof}
Let $f$ be a smooth function to be
chosen  later, then $$\int
|f^{-1}d(fu)|^2=\int|du|^2+f^{-2}|df|^2u^2+2f^{-1}u\langle df,du\rangle\geq 0$$
An integration by parts shows that
$$
\int 2f^{-1}u\langle df,du\rangle=\int u^2f^{-2}|df|^2+u^2f^{-1}\nabla^*\nabla f.
$$
This leads to
$$  \int u\nabla^*\nabla
u=\int|du|^2\geq\int(-f^{-1}\nabla^*\nabla f-2f^{-2}|df|^2)u^2\;.
$$ When $f=\rho^{-\frac{n-1}{2}}$ the last term equals
$(n-1)^2\|u(1+O(\rho))\|^2_{L^2}/4$, which concludes the proof.
\end{proof}

Lemma~\ref{Lnew} combined with the fact that
$V^{-2}|dV|^2=1+O(\rho)$ and that $ V^{-1}\nabla^*\nabla
V=-3+O(\rho)$ shows that
$$
\|u\|_{L^2}\|Zu\|_{L^2}\ge -\int uZu\geq
\Big(\frac{(3-1)^2}{4}+\frac{15}{4}-\frac{9}{2}\Big)\int(1+O(\rho))u^2,$$
which shows that $ Z$ satisfies the condition \eq{Lec1.4}
with
$$C=\Big(\frac{(3-1)^2}{4}+\frac{15}{4}-\frac{9}{2}\Big)^{-1/2}={2}\;.$$
We recall that the critical weight to be in $L^2$ is $O(\rho^{1})$ so
the function $f=V^{-3/2}=O(\rho^{3/2})$, corresponding to $\omega=1$,
is in the $L^2$-kernel of $Z$. We prove now that this kernel
equals
 $$
\ker Z=V^{\frac{-3}{2}}\R\;.
 $$
 Assume $f$ is in the $L^2$-kernel of $\mathcal Z$, by elliptic
regularity $f$ is smooth on $M$. Let $\varphi_k \in W^{1,\infty}$ be
any function on $M$ such that $\varphi_k=1$ on the geodesic ball
$B_p(k)$ of radius $k$ centred at $p$, with $\varphi_k=0$ on
$M\setminus B_p(k+1)$, and $|\nabla \varphi_k|\le C$ independently of
$k$.  Such functions can be constructed by composing the geodesic
distance from $p$ with a test function on $\R$. Integrating by parts
one has
\begin{eqnarray*}
0 & = & - \int V^3\varphi_k^2 f \cZ f=-\int  \varphi_k^2  f
\nabla^i(V^{-3}\nabla_if)\\
& = & \int  \varphi_k^2  V^{-3}|\nabla f|^2+2  V^{-3}f  \varphi_k   \nabla^i\varphi_k
\nabla_if
\end{eqnarray*}
Using H\"older's inequality, the second integral can be estimated from
below by
$$- 2\left(\int  \varphi_k^2  V^{-3}|\nabla f|^2\right)^{1/2}
\left(\int  f^2  V^{-3}|\nabla \varphi_k|^2\right)^{1/2}\;,$$
leading to
$$\int \varphi_k^2 V^{-3}|\nabla f|^2\le 4 \int f^2 V^{-3}|\nabla
\varphi_k|^2\;.$$ By Lebesgue's dominated convergence theorem, the
right-hand side converges to zero as $k$ tends to infinity because
$f\in L^2$, while $V^{-1}$ is uniformly bounded, and $\nabla
\varphi_k$ is supported in $B_p(k+1)\setminus B_p(k)$. So $f$ is a
constant.  Using \cite{Lee:fredholm}, Theorem C(c), we thus obtain

\begin{theorem}\label{isofunction3d}
Let $(V,g)$ be close in $C_{-1}^{k+2,\alpha}(M)\times
C_{0}^{k+2,\alpha}(M,{\mathcal S}_2)$ to an asymptotically
hyperbolic static metric. Then $Z$ is an isomorphism from
$C^{k+2,\alpha}_{\delta}( M)/V^{-3/2}\R$ to
$$\Big\{f\in C^{k,\alpha}_{\delta}(M): \int_M V^{-3/2}f=0\Big\}\;.$$
 when
$1/2<\delta<\frac{3}{2}$. Equivalently, ${\mathcal Z}$ is an
isomorphism from $C^{k+2,\alpha}_{\delta}( M)/\R$ to
\bel{intcond}\Big\{f\in C^{k,\alpha}_{\delta}(M): \int_M V^{-3}f=0\Big\}\;.\ee
when $-1<\delta<0$.
\end{theorem}

\section{The equations}

Rescaling the metric to achieve a convenient normalisation of the
cosmological constant, the vacuum Einstein equations for a metric
satisfying \eq{gme1}-\eq{gme2} read (see, e.g.,
\cite{Coquereaux:1988ne})
\begin{equs}\label{mainequation}
\left\{\begin{array}{l} V(\nabla^*\nabla V+n V)=\frac 1{4} |\newF|_g^2\;,\\
     \Ric(g)+n g-V^{-1}\Hess_gV=\frac{1}{2V^{2}}\newF\circ \newF\;,
\\
\div (V \newF)=0\;,
\end{array}\right.
\end{equs}
where
$$\newF_{ij}=-V^2(\partial_i \theta_j - \partial_j
\theta_i)\;,\;\;\;(\newF\circ \newF)_{ij}=\newF_i{^k}\newF_{kj}\;.$$
In dimension $n=3$ an alternative set of equations can be obtained by
introducing the \emph{twist potential} $\omega$. Writing
$d\omega=\omega_idx^i$ one sets $$\omega_i = \frac {V}2
\epsilon_{ijk}\lambda^{jk}
\quad \Longleftrightarrow \quad \lambda_{jk}=\frac 1{V}
\epsilon_{jk\ell}\omega ^\ell\;.$$
This leads to (compare~\cite{Exactsolutions})
\begin{equs}\label{mainequation3d}
\left\{\begin{array}{l} V(\nabla^*\nabla V+3V)=\frac 1{2V^{2}} |d\omega|^2\;,\\
     \Ric(g)+3g-V^{-1}\Hess_gV=\frac{1}{2V^4} (d\omega \otimes d\omega-
|d\omega|^2 g) \;,
 \\ \nabla^*
  (V^{-3}\nabla \omega)=0\;.
\end{array}\right.
\end{equs}

\subsection{The linearised equation}\label{sec:line}
We first consider the operator from the set of functions times
symmetric two tensor fields to itself, defined as
$$
\left(\begin{array}{l}V\\g\end{array}\right)\mapsto
\left(\begin{array}{l}V(\nabla^*\nabla
V+nV)\\\Ric(g)+ng-V^{-1}\Hess_g V\end{array}\right)\;.
$$ The two components of its linearisation at $(V,g)$ are
$$
p(W,h)=V\left[(\nabla^*\nabla+2n+V^{-1}\nabla^*\nabla V) W
+ \langle \Hess_gV,h\rangle_g-\langle \div\grav
h,dV\rangle_g\right]\;,
$$
\begin{eqnarray*}
P_{ij}(W,h)&=&\frac{1}{2}\Delta_Lh_{ij}+nh_{ij}+\frac{1}{2}V^{-1}\nabla^kV(\nabla_ih_{kj}+
\nabla_jh_{kj}-\nabla_kh_{ij})-(\div^*
\div\grav h)_{ij}\\
&&+V^{-2}W(\Hess_gV)_{ij}-V^{-1}(\Hess_g W)_{ij}\;.
\end{eqnarray*}

We let $\Tr$ denote the trace and we set
$$
\grav h=h-\frac{1}{2}\Tr_g hg,\;\;\; (\div
h)_i=-\nabla^kh_{ik},\;\;\;
(\div^*w)_{ij}=\frac{1}{2}(\nabla_iw_j+\nabla_jw_i)\;,$$ (note the
geometers' convention to include a minus in the definition of
divergence).
 It turns out to be convenient to introduce the one-form
$$
w_j=V^{-1}\nabla^kVh_{kj}+\nabla^kh_{kj}-\frac{1}{2}\nabla_j(\Tr
h)-V^{-1}\nabla_jW-V^{-2}\nabla_jV W\;,
$$
which allows us  to rewrite  $P(W,h)$ as 
\begin{eqnarray*}
P(W,h)
&=&L(W,h)+\div^*w\;,
\end{eqnarray*}
where $L$ is as in Corollary~\ref{isotordu}.  Similarly, $p(W,h)$ can
be rewritten as
\begin{eqnarray*}
p(W,h)
&=&l(W,h)+V\langle w,dV\rangle_g\;.
\end{eqnarray*}

\subsection{The modified equation}\label{secjauge}

We want to use the implicit function theorem to construct our
solutions. As is well known, the linearisation of the Ricci tensor
does not lead to  well behaved equations, and one adds ``gauge
fixing terms" to take care of this problem. Our  choice of those
terms arises from harmonic coordinates for the vacuum Einstein
equations in one dimension higher.

In dimension 3, we start by solving the following  system of
equations
\be\label{jaugeequation3d} \left\{\begin{array}{lll}q(V,g)&:=&
V(\nabla^*\nabla V+3V+\langle \Omega,dV\rangle)-\frac 1{2V^{2}}
|d\omega|^2=0\;,\\
Q(V,g)&:=&\Ric(g)+3g-V^{-1}\Hess_gV+\div^*\Omega\\&&\hspace{3cm}-\frac{1}{2V^4}
(d\omega d\omega- |d\omega|^2 g)=0 \;,\\ &&\nabla^* (V^{-3}\nabla
\omega)=0\;,
\end{array}\right.
\ee
with
\newcommand{\tGamma}{\w{\Gamma}}%
\newcommand{\zGamma}{\mathring\Gamma}%
\newcommand{\hGamma}{\hat\Gamma}%
\newcommand{\wg}{\hat g}%
\newcommand{\znabla}{\mathring\nabla}%
\newcommand{\zV}{U}%
\begin{eqnarray}
\nonumber
-\Omega_j&\equiv&
-\Omega(V,g,U,b)_j
\\
\nonumber
&:=&
\wg_{j\mu}\wg{}^{\alpha \beta} (\hGamma^\mu_{\alpha\beta}- \tGamma^\mu_{\alpha\beta})
\\
\nonumber
& = &
{g}_{jk}{g}{}^{\ell m} (\Gamma^k_{\ell m}- \zGamma^k_{\ell m}) +
V^{-2}g_{jk}(\zV\znabla^j \zV - V\nabla^j V)
\\
& = &
{g}{}^{\ell m} (\znabla_m {g}_{j\ell}- \frac 12 \znabla_jg_{\ell m}) + V^{-2}g_{jk}(\zV\znabla^j \zV - V\nabla^j V)
\label{Omegaeq}
\end{eqnarray}
where $\mathring\nabla$-derivatives are relative to a fixed metric
$b$ with Christoffel symbols $\zGamma^\alpha_{\beta\gamma}$, $U$ is
a fixed positive function, latin indices run from $0$ to $n$, and
$\wg:=V^2(dx^0)^2+g$ with Christoffel symbols
$\hGamma^\alpha_{\beta\gamma}$, while the
$\tGamma^\alpha_{\beta\gamma}$'s are the Christoffel symbols of the
metric $\zV^2(dx^0)^2+b$, compare~\eq{Christof} below. The co-vector
field $\Omega$ has been chosen to contain terms which cancel the
``non-elliptic terms'' in the Ricci tensor, together with some
further terms which will ensure bijectivity of the operators
involved. The second line of the equation above makes clear the
relation of $\Omega$ to the $n+1$-dimensional metric $\wg$ and its
$(\zV,b)$-equivalent.

In dimension $n$, as a first step we will solve the  system
\begin{equs}\label{jaugeequation}
\left\{\begin{array}{l} q(V,g):=V(\nabla^*\nabla V+n V+\langle \Omega,dV\rangle)-\frac 1{4} |\newF|_g^2=0
\;,\\
   Q(V,g):=  \Ric(g)+n g-V^{-1}\Hess_gV+\div^*\Omega-\frac{1}{2V^{2}}\newF\circ \newF=0\;,
\\
\div (V \newF)=-V^3d\sigma\;,
\end{array}\right.
\end{equs}
where $\Omega$ is as in dimension 3, while the ``Lorenz-gauge
fixing function'' $\sigma$ equals
 $$\sigma=V^{-3}\nabla^i(V^{3}\theta_i)\;.$$
A calculation shows
$$ \div (V \newF)+V^3d\sigma=V^3[-{\mathcal Q}+2(V^{-1}\nabla^*\nabla
V+n)](\theta)\;,
$$
 where $\mathcal Q$ is as in Corollary~\ref{isotrans}, which makes
clear the elliptic character of the third equation in
\eq{jaugeequation}.

The derivative of $\Omega$ with respect to $(V,g)$ at $(U,b)$ is $$
D_{(V,g)}\Omega(U,b)(W,h)=-w,$$ where $w$ is the one-form defined in
Section~\ref{sec:line} with $(V,g)$ replaced with $(U,b)$.  Thus, the
linearisation of $(q,Q)$ at $(U,b)$ is
$$ D(q,Q)(U,b)=(l,L)\;,$$ where $(l,L)$ is the operator defined in
Section~\ref{sec:line} with $(V,g)$ replaced with $(U,b)$.  We will
show that, under reasonable conditions, solutions of
\eq{jaugeequation3d} (resp.  \eq{jaugeequation}) are solutions of
\eq{mainequation3d} (resp.~\eq{mainequation}). If $(\omega,V,g)$
solves \eq{jaugeequation3d} (resp.  if $(\theta,V,g)$ solves
\eq{jaugeequation}), we set
$$\Phi:=\div^*\Omega\;,$$
$$
a:=\left\{\begin{array}{lll} \frac 1{2V^{4}} |d\omega|^2 &
\mbox{in the context of \eq{jaugeequation3d}},\\
\frac 1{4V^2} |\newF|_g^2& \mbox{in the context of
\eq{jaugeequation}},\end{array}\right.$$
$$
A:=\left\{\begin{array}{lll}    \frac{1}{2V^4} (d\omega d\omega-
|d\omega|^2 g)&
\mbox{in the context of \eq{jaugeequation3d}},\\
\frac{1}{2V^{2}}\newF\circ \newF& \mbox{in the context of
\eq{jaugeequation}.}
\end{array}\right.
$$
With this notation, the first two equations in both
\eq{jaugeequation3d} and \eq{jaugeequation} take the form
\begin{equs}\label{jaugeequationag}
\left\{\begin{array}{l} \nabla^*\nabla V+n V+\langle \Omega,dV\rangle=V a
\;,\\
    \Ric(g)+n g-V^{-1}\Hess_gV+\Phi=A\;.
\end{array}\right.
\end{equs}
If we take the trace of the second equation in~\eq{jaugeequationag} we
obtain
\begin{eqnarray*}
0&=&R(g)+n^2+V^{-1}\nabla^*\nabla V+\Tr\Phi-\Tr
A\\&=&R(g)+n^2-n-V^{-1}\langle \Omega(V,g),dV\rangle+\Tr\Phi+a-\Tr
A\;.
\end{eqnarray*}
 Then
\begin{eqnarray*}
 E(g)&:=&\grav_g \Ric(g)\\
&=&-ng+V^{-1}\Hess V-\Phi-\frac{1}{2}(-n(n-1)+V^{-1}\langle
\Omega(V,g),dV\rangle-\Tr\Phi)g\\
&&\;\;+\grav_g A+\frac{a}{2}g\;.\\
\end{eqnarray*}
As usual, we will use the vanishing of the divergence of $E$ to
obtain an equation for $\Omega$.  For a solution to the modified
equation, the  divergence of $E(g)$ equals
\begin{eqnarray*}
 \div E(g)_j&=&V^{-2}\nabla^iV\nabla_i\nabla_jV-V^{-1}\nabla^i\nabla_j\nabla_iV+\nabla^i\Phi_{ij}
 -\frac{1}{2}\nabla_j\Tr\Phi
\\
 &&+\frac{1}{2}\nabla_j(V^{-1}\langle
\Omega(V,g),dV\rangle)+\Big(\div(\grav_g
 A+\frac{a}{2}g)\Big)_j
\\
&=&V^{-1}\nabla^iV(R_{ij}+ng_{ij}+\Phi_{ij}-A_{ij})-V^{-1}\nabla^iVR_{ij}
\\
&&
-V^{-1}\nabla_j(nV+\langle
\Omega(V,g),dV\rangle+Va)
+\nabla^i\Phi_{ij}
 -\frac{1}{2}\nabla_j\Tr\Phi
\\
&&
 +\frac{1}{2}\nabla_j(V^{-1}\langle
\Omega(V,g),dV\rangle)+\Big(\div(\grav_g
 A+\frac{a}{2}g)\Big)_j
\\
&=&\nabla^i\Phi_{ij}+V^{-1}\nabla^iV\Phi_{ij}-\frac{1}{2}\nabla_j\Tr\Phi-\frac{1}{2}V^{-1}\nabla_j\langle
\Omega(V,g),dV\rangle
\\
&&-\frac{1}{2}V^{-2}\nabla_jV\langle
\Omega(V,g),dV\rangle
+\underbrace{V^{-1}\nabla_j(Va) - V^{-1}\nabla^i V A_{ij}
+\Big(\div(\grav_g
 A+\frac{a}{2}g)\Big)_j}_{=:\beta_j}
\\
&=&\nabla^i\Phi_{ij}+V^{-1}\nabla^iV\Phi_{ij}-\frac{1}{2}\nabla_j\Tr\Phi
-\frac{1}{2}V^{-2}\nabla_j(V\langle
\Omega(V,g),dV\rangle)
+\beta_j
\\
&=&\frac{1}{2}[\nabla^k\nabla_k\Omega_j+V^{-1}\nabla^iV\nabla_i\Omega_j
-V^{-2}\nabla_jV\nabla^iV\Omega_i+R_{ij}
\Omega^i-V^{-1}\nabla_j\nabla^iV\Omega_i]
+\beta_j
\\
&=&\frac{1}{2}[B(\Omega)_j+R_{ij}
\Omega^i-V^{-1}\nabla_j\nabla^iV\Omega_i]
+\beta_j\\
&=&\frac{1}{2}{\mathcal
B}(\Omega)_j+\beta_j\;.\\
\end{eqnarray*}
We now claim that $\beta_j$ vanishes when $\sigma$ does.  For
\eq{jaugeequation3d} this is a straightforward computation.  For
\eq{jaugeequation} we have
\begin{eqnarray*}
-\beta_j
 &=&
 \frac 12 V \newF_j{^k}\nabla_k \sigma
 -\frac 12 V^{-3}\nabla_j V |\newF|^2 +\frac 18V^{-2} \nabla_j|\newF|^2 \\&&  +\frac 12
V^{-2} \newF^{ik}\nabla_i \newF_{kj}+V^{-3}\nabla^iV
\newF_i{^k}\newF_{jk}\;.
\end{eqnarray*}
{}From the definition of $\newF_{ij}$ one has
$$
\nabla_{[i}( V^{-2}\newF_{kj]})=0\;.
$$
This gives
$$\newF^{ik}(2 \nabla_{i}\newF_{kj}+
\nabla_j \newF_{ik} - 6 V^{-1}(\nabla_{[i} V)\newF_{kj]})=0\;,
$$
which can also be rewritten as $$\newF^{ik}\nabla_i\newF_{kj} = -
\frac 14 \nabla_j |\newF|^2 + 2V^{-1} \nabla_i V
\newF^i{_k}\newF^k{_j} + V^{-1} \nabla_j V |\newF|^2\;,
$$
and our claim follows.

 We will see during the construction to follow that solutions of the
third equation in~\eq{jaugeequation} which decay sufficiently fast
satisfy $\sigma=0$. The Bianchi identity $\div E(g)=0$ shows then that
$\Omega$ is in the kernel of ${\mathcal B}$. It follows from
Corollary~\ref{cor:isoform} that the only solution of this equation
which decays sufficiently fast is zero.

\section{The construction}
\subsection{The $n$-dimensional case}
We consider ${\mathring V}^2d\varphi^2+{\mathring g}$, an
asymptotically hyperbolic Einstein static metric on $\mbbS^1\times
M$. We prescribe
 $\hat{\theta}\in C^{k+2,\alpha}(\partial M,{\mathcal
T}_1)$, and we seek a  solution
$$ \theta=\theta(\hat{\theta},V,g)\in C_{1}^{k+2,\alpha}(M,{\mathcal T}_1)
$$
of the problem
 \bel{equatheta}
 \left\{\begin{array}{l} \div (V
\newF)+V^3d\sigma\equiv
V^{3}[-{\mathcal Q}+2(V^{-1}\nabla^*\nabla V+n)]\theta=0\;,\\
\theta-\hat{\theta}\in C_{2}^{k+2,\alpha}(M,{\mathcal T}_1)\;;
\end{array}\right.
 \ee
recall that
\bel{newFag}
\newF_{ij}=-V^2(\partial_i \theta_j
-\partial_j\theta_i) \mbox{ and }
\sigma=V^{-3}\nabla^i(V^{3}\theta_i)\;.
\ee
Such solutions can be
obtained by solving the following equation for
$\theta-\hat\theta$:
 $$
 [-{\mathcal Q}+2(V^{-1}\nabla^*\nabla V+n)](\theta-\hat\theta)
 =-[-{\mathcal Q}+2(V^{-1}\nabla^*\nabla V+n)]\hat\theta\;.
 $$
When $V=\mathring V$ in \eq{equatheta}, then the term
$V^{-1}\nabla^*\nabla V+n$ vanishes and the operator is an
isomorphism by Corollary~\ref{isotrans} with $\delta = 1$. Thus,
the operator appearing there is an isomorphism for all nearby
$V$'s. In fact, for any Riemannian metric $g$ on $M$, close to
${\mathring g}$ in $C_{0}^{k+2,\alpha}(M,{\mathcal S}_2)$, with
$g-\mathring g \in C_{1}^{k+2,\alpha}(M,{\mathcal S}_2)$,  and for
any function $V$ on $M$, close to ${\mathring V}$ in
$C_{-1}^{k+2,\alpha}(M)$, with $V-V_0\in C_{0}^{k+2,\alpha}(M)$ a
unique solution exists. Moreover the map
$(\hat{\theta},V,g)\mapsto \theta-\hat{\theta}$ is smooth.

Let us denote by ${\mathring \theta}$ the solution of \eq{equatheta}
with $(V,g)=({\mathring V},{\mathring g})$.

\medskip

\begin{remark}
\label{R5.3}
${\mathring \theta}$ is polyhomogenous when ${\mathring V}$ and
${\mathring g}$ are by the results in~\cite{AndChDiss}. Applying the
second line of \eq{equatheta} twice we obtain
$$\theta-{\mathring
\theta}=\theta-\hat\theta+\hat\theta-{\mathring \theta}\in
C_{2}^{k+2,\alpha}(M,{\mathcal T}_1)\;.
$$
Furthermore, one has directly that $\sigma-{\mathring \sigma}\in
C_{2}^{k+1,\alpha}(M)$; ${\mathring \sigma}$ is in fact also in
$C_{2}^{k+1,\alpha}(M)$ by expanding near the boundary.
\end{remark}

\bigskip

Suppose that $\theta$ solves $\div (V \newF)+V^3d\sigma=0$, then
clearly $$\div [\div (V \newF)+V^3d\sigma]=0.$$ Since the double
divergence of any anti-symmetric tensor vanishes identically it holds
that $\div \div (V\lambda)=0$, so that under \eq{equatheta} $\sigma$
is in $ C_{2}^{k+1,\alpha}(M) $ and verifies
$$
\nabla^i(V^3\nabla_i\sigma)=0.$$ It follows from Theorem
\ref{isofunction} that $\sigma=0$ when $n\geq2$.

Let us define a map $F$, from one-forms on $\partial_\infty M$
times functions on $M$ times symmetric two-tensor fields on $M$ to
functions on $M$ times symmetric two-tensor fields, which to
$(\hat{\theta},V,g)$ associates $$ \left(\begin{array}{c}
V(\nabla^*\nabla V+n V+\langle \Omega(V,g,{\mathring V},{\mathring
g}),dV\rangle)-\frac 1{4} |\newF|_g^2\;\\ \Ric(g)+n
g-V^{-1}\Hess_gV+\div^*\Omega(V,g,{\mathring V},{\mathring
g})+\frac{1}{2V^{2}}\newF\circ
\newF
\end{array}\right)\;.$$

\begin{proposition}\label{Flisse}
Let ${\mathring V}^2d\varphi^2+{\mathring g}$ be an asymptotically
hyperbolic static Einstein metric on $\mbbS^1\times M$, $k\in\N$,
$\alpha\in(0,1)$. The map ${\mathcal   F}$ defined as
$$
\begin{array}{ccc}
C^{k+2,\alpha}(\partial M,{\mathcal T}_1)\times
C_{1}^{k+2,\alpha}(M)\times C_{2}^{k+2,\alpha}(M,{\mathcal
S}_2)&\longrightarrow&C_{0}^{k,\alpha}(M)\times
C_{2}^{k,\alpha}(M,{\mathcal  S}_2)\\
(\hat{\theta},W,h)&\longmapsto&F(\hat{\theta},{\mathring
V}+W,{\mathring g}+h)
\end{array}
$$ is smooth in a neighborhood of zero.
\end{proposition}
\begin{proof}
The function ${\mathring V}\in C_{-1}^{k+2,\alpha}(M)$ is strictly
positive, so the same is true for ${\mathring V}+W$ if $W$ is
sufficiently small in $C_{1}^{k+2,\alpha}(M)\subset
C_{-1}^{k+2,\alpha}(M) $. Similarly, the symmetric two-tensor
field ${\mathring g}+h\in C_{0}^{k+2,\alpha}(M,{\mathcal S}_2) $
is positive definite when $h$ is small in
$C_{2}^{k+2,\alpha}(M,{\mathcal S}_2)\subset
C_{0}^{k+2,\alpha}(M,{\mathcal S}_2)$. The map
$(\hat{\theta},V,g)\mapsto \theta$ is smooth. Now, for $\theta\in
C_{1}^{k+2,\alpha}(M)$, by \eq{newFag} and by Remark~\ref{R5.3} we
have
$$
\lambda_{ij}=O(\rho^{-2})\;,
$$
which further implies
$$
\frac{1}{2V^{2}}\newF\circ \newF\in C_{2}^{k,\alpha}(M,{\mathcal
S}_2)\;.
$$ The fact that the remaining terms in $F(\hat{\theta},{\mathring
V}+v,{\mathring g}+h)$ are in the space claimed, and that the map is
smooth is standard (see \cite{Lee:fredholm} for instance).
\end{proof}

We can conclude now as follows:

\begin{theorem}\label{imp}
Let $n\ge 2$, and let ${\mathring V}^2d\varphi^2+{\mathring g}$ be a
polyhomogenous non-degenerate asymptotically hyperbolic static
Einstein metric on $\mbbS^1\times M$, $k\in\N\backslash\{0\}$,
$\alpha\in(0,1)$. For all $\hat{\theta}$ close to zero in
$C^{k+2,\alpha}(\partial M,{\mathcal T}_1)$, there exists a unique
solution
$$(\theta,V,g)=({\mathring \theta}+\vartheta,{\mathring
V}+W,{\mathring g}+h)$$ to \eq{mainequation} with ${\mathring
\theta}-\hat{\theta}\in C_{2}^{k+2,\alpha}(M)$ and
$$
(\vartheta,W,h)\in C_{3}^{k+2,\alpha}(M)\times
C_{1}^{k+2,\alpha}(M)\times C_{2}^{k+2,\alpha}(M,{\mathcal
S}_2)\;,
$$ close to zero, satisfying the gauge conditions
$\Omega=\sigma=0$. Moreover, the maps $\hat{\theta}\mapsto {\mathring
\theta}-\hat{\theta}$ and $\hat{\theta}\mapsto (\vartheta,W,h)$ are
smooth maps of Banach spaces near zero.
\end{theorem}

\begin{proof}
As already pointed out, the one-form
$\theta=\theta(\hat{\theta},V,g)$ exists and is
unique when $W$ and $h$ are small. From Proposition \ref{Flisse} we
know that the map ${\mathcal F}$ is smooth. The linearisation of
${\mathcal F}$ at zero is
$$
D_{(W,h)}{\mathcal   F}(0,0,0)=D_{(V,g)}F(0,{\mathring
V},{\mathring g})=(l,L)\;.
$$
From Corollary~\ref{isotordu}, with $\delta=2$, we obtain that
$D_{(W,h)}{\mathcal F}(0,0,0)$ is an isomorphism. The implicit
function theorem shows that the conclusion of Theorem~\ref{imp}
remains valid \emph{for the modified equation~\eq{jaugeequation}}.
Returning to Section~\ref{secjauge}, we see that
$\Omega=\Omega(V,g,{\mathring V},{\mathring g})\in
C_{2}^{k+1,\alpha}(M,T_1)$ and that ${\mathcal B}(\Omega)=0$, so
from Corollary \ref{cor:isoform}, we have $\Omega=0$, obtaining
thus a solution to \eq{mainequation}.
\end{proof}

\subsection{The three-dimensional case}

In three dimensions an alternative construction can be given, as
follows.  We consider again an asymptotically hyperbolic Einstein
static metric ${\mathring V}^2d\varphi^2+{\mathring g}$ on
$\mbbS^1\times M$. We use Theorem \ref{isofunction3d}, with $g$
--- a Riemannian metric on $M$ close to ${\mathring g}$ in
$C_{0}^{k+2,\alpha}(M,{\mathcal S}_2)$, and $V$ --- a function on $M$
close to ${\mathring V}$ in $C_{-1}^{k+2,\alpha}(M)$. For our purposes
there is no preferred value of parameter $\delta$ there. It is
convenient to set $\delta=s-1$, and arbitrarily choose some
$s\in(0,1)$. For any $\hat{\omega}\in C^{k+1,\alpha}(\partial M)$
satisfying
\bel{bintc} \int_{\partial M} \hat \omega =0\;,
\ee
there
exists a unique, modulo constant, solution
$$ \omega=\omega(\hat{\omega},V,g)\in C_{-1}^{k+2,\alpha}(M)$$ to
$$
\left\{\begin{array}{l} \nabla^*(V^{-3}\nabla\omega)=0,\\
\omega-\hat{\omega}\rho^{-1}\in C_{s-1}^{k+2,\alpha}(M)\;.
\end{array}\right.
$$ (This can be seen by writing ${\mathcal Z}\delta \omega =
-{\mathcal Z}(\hat \omega\rho^{-1})$, and checking that the source
term in the equation for $\delta \omega$ satisfies the
integrability condition \eq{intcond} when \eq{bintc} holds.)
 Moreover, the map $(\hat{\omega},V,g)\mapsto
\omega-\hat{\omega}\rho^{-1}$ is smooth in the
$C_{s-1}^{k+2,\alpha}(M)$ topology.

We define a new map $F$, defined on the set of functions on
$\partial_\infty M$ times functions on $M$ times symmetric two-tensor
fields, mapping to functions on $M$ times symmetric two-tensor fields,
which to $(\hat{\omega},V,g)$ associates
$$
\left(\begin{array}{c}V(\nabla^*\nabla V+3V+\langle
\Omega(V,g,{\mathring V},{\mathring g}),dV\rangle)-\frac 1{2V^{2}} |d\omega|^2\\
\Ric(g)+3g-V^{-1}\nabla_i\nabla_jV+\div^*\Omega(V,g,{\mathring
V},{\mathring g})-\frac{1}{2V^4} (d\omega d\omega- |d\omega|^2
g)\end{array}\right)\;.
$$

\begin{proposition}\label{Flisse3d}
Let ${\mathring V}^2d\varphi^2+{\mathring g}$ be an asymptotically
hyperbolic static Einstein metric on $\mbbS^1\times M$, $k\in\N$,
$\alpha\in(0,1)$. The map ${\mathcal   F}$ defined as
$$
\begin{array}{ccc}
C^{k+2,\alpha}(\partial M)\times C_{1}^{k+2,\alpha}(M)\times
C_{2}^{k+2,\alpha}(M,{\mathcal
S}_2)&\longrightarrow&C_{0}^{k,\alpha}(M)\times
C_{2}^{k,\alpha}(M,{\mathcal  S}_2)\\
(\hat{\omega},W,h)&\longmapsto&F(\hat{\omega},{\mathring
V}+W,{\mathring g}+h)
\end{array}
$$
is  smooth  in a neighborhood of zero.
\end{proposition}

\begin{proof}
The proof is essentialy the same as that of Proposition~\ref{Flisse}. We simply note that
for all $s\in (0,1)$ we have, by direct estimations,
$$
V^{-4} (d\omega d\omega- |d\omega|^2 g)\in
C_{2}^{k,\alpha}(M,{\mathcal S}_2)\;.
$$
\end{proof}

We are ready to  formulate now:

\begin{theorem}\label{imp3d}
Let $\dim M=3$ and let ${\mathring V}^2d\varphi^2+{\mathring g}$ be a
non-degenerate asymptotically hyperbolic static Einstein metric on
$\mbbS^1\times M$, $k\in\N$, $\alpha\in(0,1)$, $s\in(0,1)$. For all
$\hat{\omega}$ close to zero in $C^{k+2,\alpha}(\partial M)$ and
satisfying \eq{bintc} there exists a unique solution
$$(\omega,V,g)=(\hat{\omega}\rho^{-1}+w,{\mathring V}+W,{\mathring
g}+h)$$ to \eq{mainequation3d} with
$$
(w,W,h)\in C_{s-1}^{k+2,\alpha}(M)\times
C_{1}^{k+2,\alpha}(M)\times C_{2}^{k+2,\alpha}(M,{\mathcal
S}_2)\;,
$$ close to zero, satisfying the gauge condition $\Omega=0$. Moreover,
the map $\hat{\omega}\mapsto (w,W,h)$ is a smooth map of Banach spaces
near zero.
\end{theorem}

\begin{proof}
The proof is identical to that of Theorem~\ref{imp}, making use of
Proposition~\ref{Flisse3d}.
\end{proof}

\section{Uniqueness}
\label{SecUni}

So far we have shown that solutions are unique in the gauge
$\Omega=0$, together with the condition $\sigma=0$ in the context of
\eq{mainequation}.  We claim that any metrics satisfying the
hypotheses of Theorem~\ref{maintheorem} can be brought to this gauge.

First, consider $\sigma$; note that the one-form $\theta$ of \eq{gme1}
is defined modulo the differential of a function $f$ defined on $M$;
indeed, the replacement $t\to t+f$ leads to $\theta\to \theta+df$.  We
can then use Theorem~\ref{isofunction} to find a unique $f$ such that
the function $\sigma$ associated with $\theta+df$ vanishes.

The vanishing of $\Omega$ requires a smallness hypothesis, as well as
some work. Suppose that we are given a couple $(V,g)$ near to
$(\mathring V,\mathring g)$. The second line of \eq{Omegaeq} shows
that, in the notation of~\cite{CDLS}, the condition $\Omega=0$ is
exactly the condition $\Delta_{g'\w g} \Id =0$, where $g'=
V^2d\varphi^2 + g$. The proof that $\Omega$ can be made to vanish is
established by inspection of the arguments of Section~4
of~\cite{CDLS}. We simply note that the implicit function theorem, as
invoked there, can be applied globally on $M$ (rather than in a collar
neighborhood of the boundary, as in~\cite{CDLS}) if we assume that
$(\mathring V/V,V^{-2}g)$ is close to $(1,\mathring V^{-2}\mathring
g)$ in $C^{2}(\overline M)$. Indeed, the linearised operator, denoted
by $L$ in~\cite{CDLS}, is again an isomorphism by the results
of~\cite{Lee:fredholm}, as follows from the fact that $\tilde g$ is
Einstein, with negative scalar curvature. (Actually, the Einstein
equations are irrelevant for the question of $\Omega=0$ gauge, as long
as the Ricci tensor of $\tilde g$ is negative definite.)

Uniqueness of solutions up-to-diffeomorphism (which is the identity on
the boundary) is a direct consequence of the above.




\section{Polyhomogeneity}\label{section:polyhomogeneity}

Let $U_0\subset \R^n$ be an open set, and let $U=
U_0\cross(0,\epsilon) $ with coordinates $(x,y)$.  For any
$\delta\in\R$, we denote by $\scr C^\delta$ the space of functions
$f\in C^\infty(U)$ that satisfy, on any subset
$K\cross(0,\epsilon_0)$ with $K\subset U_0$ compact and
$0<\epsilon_0<\epsilon$, estimates of the following form for all
integers $r\ge 0$ and all multi-indices $\alpha$:
\begin{displaymath}
\left| (y\del_y)^r \del_ x^\alpha f(x,y)\right| \le C_{r,\alpha}
y^\delta.
\end{displaymath}
(We use the multi-index notations $\alpha =
(\alpha_1,\dots,\alpha_{n})$ and $\del_ x^\alpha = (\del_{
x^1})^{\alpha_1}\dots (\del_{ x^n})^{\alpha_n}$.)

A smooth function $f\colon U\to \R$ is said to be {\it
polyhomogeneous} (cf.\/~\cite{mazzeo,AndChDiss}) if there exists a
sequence of real numbers $s_i \to +\infty$, a sequence of
nonnegative integers $\{q_i\}$, and  functions $f_{ij}\in
C^\infty(U_0)$  such that
\begin{equation}\label{eq:def-polyhomo}
f(x,y) \sim \sum_{i=1}^\infty \sum_{j=0} ^{q_i} y^{s_i} (\log y)^j
f_{ij}(x)
\end{equation}
in the sense that for any $\delta>0$, there exists a positive
integer $N$ such that
\begin{displaymath}
f(x,y) - \sum_{i=1}^{N}\sum_{j=0}^{q_i} y^{s_i} (\log y)^j
f_{ij}(x) \in \scr C^\delta.
\end{displaymath}
A function or tensor field on $M$ is said to be polyhomogeneous if its
coordinate representation in local coordinates near the conformal
boundary is polyhomogeneous. (We refer the reader to~\cite{ChLeski}
for a discussion of equivalence of alternative definitions of
polyhomogeneity.)

In this section, we apply the theory of \cite{AndChDiss} to conclude
that solutions to \eqref{jaugeequation} are polyhomogeneous. The key
step in the proof is a regularity result for the linearised operator
$(l,L,{\mathcal L})$. Following \cite{AndChDiss}, we say that an
interval $(\delta_-,\delta_+)\subset\R$ is a {\it weak regularity
interval} for a second-order linear operator $P$ on the spaces
$C^{k,\lambda}_\delta(M_R;S_2)$ if whenever $u$ is a locally $C^2$
section of $S_2$ such that $u\in C^{0,0}_{\delta_0}(M_R;S_2)$ and
$Pu\in C^{0,\lambda}_{\delta}(M_R;S_2)$ with $\lambda\in (0,1)$ and
$\delta_-<\delta_0<\delta<\delta_+$, it follows that $u\in
C^{2,\lambda}_{\delta}(M_R;S_2)$. We use the notation
of~\cite{CDLS}.

\begin{theorem}\label{Tphg}
Solutions given by Theorems~\ref{imp} and \ref{imp3d} are
polyhomogeneous.  Similarly, solutions of \eq{mainequation} and
\eq{mainequation3d} with smooth boundary data such that $\theta$
and $ \rho^{2}(V^2d\varphi^2+g)$ are in $ C^2(\overline M)$ are
polyhomogeneous.
\end{theorem}

\begin{proof} We start
 by noting that metrics such that $\theta$ and $
\rho^{2}(V^2d\varphi^2+g)$ are in $ C^2(\overline M)$ can be
brought, near the boundary, to a gauge in which the equations are
elliptic by setting $\sigma$ to zero as in Section~\ref{SecUni},
and then applying the results of Section~4 of~\cite{CDLS} to the
metric $V^2d\varphi^2+g$. On the other hand, solutions given by
Theorems~\ref{imp} and \ref{imp3d} are directly in the closely
related gauge $\Omega=0$; those two gauges do not coincide, but
the proof works in both gauges. Alternatively one could use the
analysis in Section~4 of~\cite{CDLS} to transform a
$C^2$--compactifiable $V^2d\varphi^2+g$ to the gauge $\Omega=0$.
A polyhomogeneous approximate solution
$\mathring{V}^{-2}(dt+\mathring \theta)^2 + \mathring g$ can then
be constructed using a Fefferman-Graham expansion
up-to-not-including the critical exponent.

For any
$\phi=(^0\phi,^2\phi,^1\phi)$, function, two-tensor, one-form on $M$,
define
\begin{displaymath}
F[\phi]:= (\rho {\mathring V}^{-1}q,\rho^2Q,{\mathcal
Q})({\mathring V}+\rho^{-1}\;^0\phi,{\mathring
g}+\rho^{-2}\;^2\phi,{\mathring \theta}+\;^1\phi)
\end{displaymath}
with $(q,Q,{\mathcal Q})$ as in \eqref{jaugeequation}, while for the
solutions arising from Theorems~\ref{imp} and \ref{imp3d} the
one-form ${\mathring \theta}$ can be taken as the solution of the
third equation in \eq{jaugeequation} with $V={\mathring V}$ and
$g={\mathring g}$.  ($F$ should not be confused with the map $F$ of
the previous section.) Then $\phi$ satisfies $F[\phi] = 0$. One can
apply \cite[Theorem 5.1.1]{AndChDiss} to $F$, and thereby conclude
that $\phi$ is polyhomogeneous. The argument proceeds as in
\cite[Section~5]{CDLS} and will not be repeated here. We simply
mention that the property, that {\it the interval $(0,n)$ is a weak
regularity interval for the operator $F'[\phi_0]$ on the spaces
$AC^\delta_{k+\lambda}(M_R)$}, is an immediate consequence of
Corollaries~\ref{isotordu} and \ref{isotrans}.
\end{proof}

\appendix
\section{``Dimensional reduction'' of some operators}
\subsection{Lichnerowicz Laplacian on two-tensor for a warped product metric}
We shall use the following coordinate systems on $S^1\times M$:
$$(x^a)=(\varphi,x^i)=(x^0,x^i)=(x^0,...,x^n).$$
\begin{lemma}\label{deltatordu}
Let $(M,g)$ be a Riemannian manifold, let $V,W$ be two functions on
$M$, let $h$ be a symmetric covariant two-tensor on $M$ and let
$\xi$ be a one-form on $M$.  On $S^1\times M$ we consider the
Riemannian metric $\w{g}=V^2d\varphi^2+g$ and the symmetric
covariant two-tensor
 $$\w{h}=2VWd\varphi^2+2\xi_id\varphi dx^i+h_{ij}dx^idx^j$$
satisfying $\mcL_{\partial_\varphi}\w{h}=0$, where $\mcL$ denotes a Lie
 derivative. Then, in local coordinates, the Laplacian of
 $\w{h}$ has the following components:
 \begin{eqnarray*}
 \w{\nabla}^c\w{\nabla}_c\w{h}_{00}&=&2[V\nabla^k\nabla_kW-\nabla^k\nabla_kVW
 -\nabla^kV\nabla_kW-V^{-1}|dV|^2W+\nabla^kV\nabla^lVh_{kl}]\;,\\
 \w{\nabla}^c\w{\nabla}_c\w{h}_{i0}&=&\nabla^k\nabla_k\xi_i-V^{-1}\nabla^k\nabla_k
V\xi_i
 -V^{-1}\nabla^kV\nabla_k\xi_i-3V^{-2}\nabla_i V\nabla^k V
 \xi_k\;,\\
\w{\nabla}^c\w{\nabla}_c\w{h}_{ij}&=&
\nabla^k\nabla_kh_{ij}+V^{-1}\nabla^kV\nabla_kh_{ij}-V^{-2}(\nabla_iV\nabla^kVh_{kj}+\nabla_jV\nabla^kVh_{ki})\\
&&+4V^{-3}\nabla_iV\nabla_jVW\;.
\end{eqnarray*}
\end{lemma}

\begin{proof}
The Christoffel symbols of the metric $\w{g}=V^2d\varphi^2+g$ are
\bel{Christof}
\w{\Gamma}^0_{00}=\w{\Gamma}^0_{ij}=\w{\Gamma}^k_{0j}=0\;,\;\;
\w{\Gamma}^k_{ij}={\Gamma}^k_{ij}\;,\;\;
\w{\Gamma}^0_{i0}=V^{-1}\nabla_i V\;,\;\;
\w{\Gamma}^k_{00}=-V\nabla^k V\;.
\ee
 The covariant derivatives of
$\w{h}$, in local coordinates,
read
$$
\begin{array}{l}\w{\nabla}_0\w{h}_{00}=2V\nabla^kV\xi_k\;,\\
\w{\nabla}_0\w{h}_{ij}=-V^{-1}(\nabla_iV\xi_j+\nabla_j V
\xi_i)\;,\\
\w{\nabla}_k\w{h}_{i0}=\nabla_k\xi_i-V^{-1}\nabla_kV\xi_i\;\\
\w{\nabla}_0\w{h}_{i0}=V\nabla^kVh_{ki}-2\nabla_iVW\;,\\
\w{\nabla}_k\w{h}_{00}=2V\nabla_kW-2\nabla_kV W\;,\\
\w{\nabla}_k\w{h}_{ij}={\nabla}_k{h}_{ij}\;.\\
\end{array}
$$
The result is obtained by substition. \end{proof}

We recall that the Lichnerowicz Laplacian is
\bel{LL}
\w{\Delta}_L\w{h}_{ab}=-\w{\nabla}^c\w{\nabla}_ch_{ab}+\w{R}_{ac}\w{h}^c{}_b
+\w{R}_{bc}\w{h}^c{}_a-2\w{R}_{acbd}\w{h}^{cd}\;.
 \ee
The curvature tensor of the warped product metric
$\widetilde{g}=V^2d\varphi^2+g$ has the following
components~\cite[Prop.~42, Chap.~7]{ONeill} (note, however, that our
curvature tensor is the negative of the one in~\cite{ONeill}):
$$
\widetilde{R}^l{}_{ijk}={R}^l{}_{ijk}\;,\quad\widetilde{R}^l{}_{0j0}=
-V\nabla_j\nabla^l
V\;,
\quad \widetilde{R}^0{}_{ijk}=
0\;,$$
$$
\widetilde{R}_{ik}={R}_{ik}-V^{-1}\nabla_k\nabla_iV,\;\;\widetilde{R}_{0k}=0,\;\;\widetilde{R}_{00}=-
V\nabla^i\nabla_iV.$$
The zero order terms in \eq{LL} are thus
$$
\begin{array}{rcl}\w{R}_{0c}\w{h}^c{}_0
&=&2\nabla^*\nabla VW\;,\\
2\w{R}_{0c0d}\w{h}^{cd}&=&-2V\nabla^i\nabla^jVh_{ij}\;,\\
\w{R}_{ic}\w{h}^c{}_0+\w{R}_{0c}\w{h}^c{}_i&=&R^l{}_i\xi_l-V^{-1}\nabla_i\nabla^lV\xi_l+V^{-1}\nabla^*\nabla
V\xi_i\;,\\
2\w{R}_{ic0d}\w{h}^{cd}&=&2V^{-1}\nabla_i\nabla_jV\xi^j\;,\\
\w{R}_{ic}\w{h}^c{}_j+\w{R}_{jc}\w{h}^c{}_i&=&R_{ik}h^k{}_j+R_{jk}h^k{}_i-V^{-1}\nabla_i\nabla_kVh^k{}_j
-V^{-1}\nabla_j\nabla_kVh^k{}_i\;,\\
2\w{R}_{icjd}\w{h}^{cd}&=&2R_{ikjl}h^{kl}-4V^{-2}\nabla_i\nabla_jVW\;.
\end{array}$$ Lemma~\ref{deltatordu} implies now:
\begin{lemma}\label{lichnetordu}
Under the hypotheses of Lemma~\ref{deltatordu}, the Lichnerowicz
Laplacian of $\w{h}$ is
  \begin{eqnarray*}
 \w{\Delta}_L\w{h}_{00}&=&2\left[-V\nabla^k\nabla_kW-\nabla^k\nabla_kVW
 +\nabla^kV\nabla_kW+V^{-1}|dV|^2W\right.
\\&&
\left.-\nabla^kV\nabla^lVh_{kl}+V\nabla^i\nabla^jVh_{ij}\right]\;,\\
\w{\Delta}_L\w{h}_{i0}&=&-\nabla^k\nabla_k\xi_i+V^{-1}\nabla^kV\nabla_k
\xi_i
 +3V^{-2}\nabla_i V\nabla^k V
 \xi_k\\&&
 +R^l{}_i\xi_l-3V^{-1}\nabla_i\nabla_jV\xi^j\;,\\
\w{\Delta}_L\w{h}_{ij}&=&
\Delta_Lh_{ij}-V^{-1}\nabla^kV\nabla_kh_{ij}+V^{-2}(\nabla_iV\nabla^kVh_{kj}+\nabla_jV\nabla^kVh_{ki})\\
&&-4V^{-3}\nabla_iV\nabla_jVW-V^{-1}(\nabla_i\nabla^kVh_{kj}+\nabla_j\nabla^kVh_{ki})\\&&+4V^{-2}\nabla_i\nabla_jVW\;.
\end{eqnarray*}
\qed
\end{lemma}

\subsection{The Laplacian on one-forms for a warped product metric}
\begin{lemma}\label{deltaform}
Let $(M,g)$ be a Riemannian manifold, let $V,f$ be two functions on
$M$ and let $\Omega$ be a one-form on $M$.  Let us consider, on
$S^1\times M$, the Riemannian metric $\w{g}=V^2d\varphi^2+g$ and the
one-form $$\w{\Omega}=fd\varphi+\Omega_idx^i\;.$$ Then in local
coordinates, the Laplacian of $\w{\Omega}$ equals $$
\w{\nabla}^c\w{\nabla}_c\w{\Omega}_{0}=\nabla^k\nabla_kf
-V^{-1}f\nabla^k\nabla_kV-V^{-1}\nabla^kV\nabla_kf,
$$
$$
\w{\nabla}^c\w{\nabla}_c\w{\Omega}_{i}=\nabla^k\nabla_k\Omega_i+V^{-1}\nabla^kV\nabla_k\Omega_i
-V^{-2}\nabla_iV\nabla^kV\Omega_k=:B(\Omega)_i\;.$$
\end{lemma}

\begin{proof}
 We have
$$
\begin{array}{l}
\w{\nabla}_0\w{\Omega}_0=V\nabla^kV\Omega_k\;,\quad
\w{\nabla}_0\w{\Omega}_i=-V^{-1}\nabla_iV f\;\\
\w{\nabla}_i\w{\Omega}_0=\partial_if-V^{-1}\nabla_iVf\;,\quad
\w{\nabla}_j\w{\Omega}_i=\nabla_j\Omega_i\;,\\
\end{array}$$
 and the result easily follows.
\end{proof}

\bigskip

\noindent{\sc Acknowledgements} The research of PTC was supported in
part by a Polish Research Committee grant 2 P03B 073 24. We are
grateful to the Isaac Newton Institute, Cambridge, for hospitality and
financial support during part of work on this paper.

\bibliographystyle{amsplain}

\bibliography{../references/newbiblio,%
../references/reffile,%
../references/bibl,%
../references/hip_bib,%
../references/newbib,%
../references/PDE,%
../references/netbiblio}
\end{document}